# Collaboration in an Open Data eScience: A Case Study of Sloan Digital Sky Survey

Jian Zhang
Drexel University
3141 Chestnut Street
Philadelphia, PA 19104
001-215-895-2474

jz85@drexel.edu

Chaomei Chen
Drexel University
3141 Chestnut Street
Philadelphia, PA 19104
001-215-895-2474

cc345@drexel.edu

## ABSTRACT

Current science and technology has produced more and more publically accessible scientific data. However, little is known about how the open data trend impacts a scientific community, specifically in terms of its collaboration behavior. This paper aims to enhance our understanding of the dynamics of scientific collaboration in the open data eScience environment via a case study of co-author networks of an active and highly cited open data project, called Sloan Digital Sky Survey. We visualized the co-authoring networks and measured their properties over time at three levels: author, institution, and country levels. We compared these measurements to a random network model and also compared results cross the three levels. The study found that 1) the collaboration networks of the SDSS community transformed from random networks to small-world networks; 2) the number of author-level collaboration instances has not changed much over time, while the number of collaboration instances at the other two levels have increased over time; 3) pairwise institutional collaboration become common in recent years. The open data trend may have both positive and negative impacts on scientific collaboration.

## Categories and Subject Descriptors

G.2.2. [**Graph Theory**]: Network problems

## 1. INTRODUCTION

Current science and technology has produced more and more publically accessible scientific data[8]. Many scientific projects primarily aim to collect scientific data, such as Sloan Digital Sky Survey (SDSS), Large Synoptic Survey Telescope (LSST)[1], and Ocean Observatory Initiative (OOI)[2]. These valuable scientific data are widely accessible through an eScience infrastructure to not only the targeted scientific communities, but other disciplines and the general public as well.

The impact of this open data trend on scientific research in general and scientific collaboration in particular, has yet been widely studied, however. Openly accessible data may build up a common ground for scientists from different institutions, disciplines, and countries. Therefore it is possible to boost scientific collaboration. On the other hand, open data could lead to competitions for publishing the first discoveries, hence perhaps hindering collaboration. This paper aims to enhance our understanding of the dynamic patterns of scientific collaboration in the open data eScience environment, and characterize the impact of open data on science collaboration. As it is believed that collaboration can promote research activity, productivity, and impact[17], knowledge about these patterns may help future scientific projects, funding agencies, and scientists to foster and benefit from collaborations.

One way to define the existence of collaboration is through co-authoring relations found in scientific publications[9]. It has long been realized that co-authorship of scientific articles provides an informative unit of analysis for studying patterns of collaboration in scientific communities[25]. In this study, we adopted this perspective and investigated scientific collaboration in the open data eScience environment via a case study of co-authorship networks of Sloan Digital Sky Survey (SDSS), which is a highly cited open data project[8,28]. The SDSS survey is one of the largest digital sky surveys up date. It collects multiple types of data of stars, galaxies, quasars, and other astronomical objects in the universe. From 2000 to the date, the SDSS has produced 30TB astronomical data, and released these data to the scientific communities and the general public through SkyServer website[3]. These data have become a real gold mine for various scientific communities and good educational materials for the general public.

---

[1] www.lsst.org
[2] www.oceanleadership.org
[3] skyserver.sdss.org

A comprehensive study of the impact of the open data trend on scientific collaboration would require comparisons between open data and non-open data projects or domains. This study mainly focuses on one particular open data project. We believe that as the first step to the study of the emerging open data phenomenon, painting a full spectrum of collaboration patterns in one project could lay down the foundation for future comparative studies in that the insights found in one project would help to generate and refine additional hypotheses and research questions for future studies.

The rest of the paper is organized as follows. Section two reviews the related work, and section three reports the methods and data used in this study. Section four presents the results of this study, while section five analyzes the most interesting results. Section six gives the conclusions of this study.

## 2. RELATED WORK

Co-authorship has long been used to study scientific collaboration. It can be traced back to the 1960s when Price and Beaver[27] used co-author relations to investigate social structures and influences in scientific communication network. They concluded that the research front of a scientific domain is dominated by a small core of active researchers and a large weak transient population of their collaborators. In later years, Beaver and Rosen systematically explored co-authorship in a series of papers[4,5,6]. And various research communities use co-author networks to map research teams and collaboration structures. For instance, [21] mapped the research departments at two universities; and [26] focused on a chemical department's collaboration.

Collaboration based on co-authorship also could be aggregated at different levels of granularity, such as institutional, interdisciplinary, topical, or international collaboration, in which two entities are considered to be in collaboration if scientists from the two entities have co-authored one or more publications. Studies[15,20] of these aggregated levels of collaboration endeavored to understand the collaborative behavior cross institutions, disciplines, and countries. Cummings and Kiesler found that multi-university collaboration created more problems than multidisciplinary collaboration[15]. A study of publications in high-energy physics found that although computer-mediated communication was believed to boost the intercollegiate and international collaboration, the percentage of papers having intercollegiate and international authorship increases evolutionary, rather than revolutionary[20].

Co-authorship can be easily transferred into a network structure, where nodes in the network represent authors, or other entities like institutions and countries, and edges represent co-authorship relations. A co-author network is one kind of graphs, various graph properties such as betweenness, bridge, centrality, clustering coefficient, degree, path length, and structure hole, could be measured to identify key nodes and edges. Exemplar studies include Chen's study of betweenness centrality of nodes to identify pivotal points in the evolution of scientific co-authors network[10], and Heinze and Bauer's study of structure holes in nano science and technology field to identify the brokerage role played by highly creative scientists in co-authorship networks.

In recent years, several social network models were applied to characterize scientific collaboration networks, such as the Erdos-Renyi model (random graphs[16]), the small-world model[30], and the scale-free model[1]. Social network models effectively describe topological characteristics across a wide range of large co-author networks in different disciplines such as biomedicine, high energy physics, astrophysics, mathematics, and computer science[22,24,25], neuroscience and mathematics[2], computer science[18], and condensed matter[9]. Newman[22,23] found that co-author networks all have a generic feature of a small-world network: a surprisingly short average distance ($L$) and a large clustering coefficient ($C$), much larger than the one expected from a random network with a similar number of nodes and edges. Other studies confirmed this observation and found the value of $L$ and $C$ varied from discipline to discipline and from database to database.

While most of the social network model studies focused on a static overview of a collaboration network within a certain time frame, some studies looked at the dynamics of structural patterns of scientific collaboration networks over time[2,7,9,18]. Both Barabasi et al, and Huang used a "snowball sampling" approach to aggregate the publications for study at different time points (mainly in year). For example given a time point $T$ between $T_{start}$ and $T_{end}$, the publications used to construct a co-author network would be all the publications from $T_{start}$ to the time point of $T$. And their studies found that while the number of authors and co-author relations kept increasing, the average distance and clustering coefficient kept decreasing in neuro science and mathematics database. Cardillo et al, directly divided publications in condense matter into one year time piece. They found the average distance slightly increased from 3.18 to 3.62 from 2000 to 2005, while the clustering coefficient is nearly constant around 0.71 throughout the six years.

A co-authorship network, to some extent, could be considered as a knowledge diffusion network in that conducting research and writing a paper is a knowledge exchanging and sharing process. Particularly, it could be a good strategy for developing institutions and countries to gain new knowledge via collaboration with advanced institutions and countries[29]. The structure of a small-world network is believed to be more efficient than that of a random network in terms of knowledge diffusion[14]. Morone and Taylor, however, found that the efficiency of knowledge diffusion in small-world network depends on the initial "knowledge gap" among network members.

In summary, a co-author network can reveal structural patterns of scientific collaboration. Network topological analysis provides information for understanding the structure of collaboration networks as well as certain information for understanding knowledge diffusion. Our research aims to utilize these methods to reveal the dynamic patterns of scientific collaboration in the SDSS publications at different aggregated levels and improve our understanding of how the open data trend impacts scientific collaboration.

## 3. METHODS

Our method consists of three steps, including data acquisition and cleanup, co-author network generation, and network analysis.

## 3.1 Dataset

The SDSS literature dataset was collected from Thomson ISI's *Web of Science* (WoS). The data were retrieved with search terms: 'SDSS' OR 'Sloan Digita*' over a time span between 2001 and 2008. A total of 2252 records were retrieved. Since the WoS has a multidisciplinary coverage, the dataset may include some records that are not relevant to the SDSS project. The abbreviation SDSS has been used for terms other than the Sloan Digital Sky Survey, for example, Strategic Decision Supporting System. In these cases, we manually removed these irrelevant records by using functions in the WoS, such as analyzing the "Document Types" and "Sources Titles". We removed records identified as "Corrections," "Letters," "Editorial materials," and "Meeting abstract." We also removed records from 51 journals and conference proceedings that clearly have nothing to do with the SDSS survey, like Water Resource Management, Diabetologia, and Cancer. The final data collection includes 2138 bibliographic records of papers.

To better identify the entities, we retrieved the metadata of each record, including the authors, their affiliations, and the countries where these affiliations are located, and compared each pair of these entities to avoid inconsistency in the data. For example, some authors put the "Los Alamos Natl Lab*s*" for "Los Alamos Natl Lab", while some authors misspell "Apache Point Observ" with "Apac*ha* Point Obser*v*" (mismatch is in italic for readability).

The author name ambiguity is a long existing problem. Recently, some ID systems, such as ResearchID at Thomson Reuter and OpenID at OpenID Foundation, have been proposed to help solve this problem. Our dataset, however, does not contain this information. Therefore, we barely used the combination of full last name and first name initials as authors' identifier. In terms of institutions and countries, we applied a levenshtein distance measurement[19] to compare two strings of affiliations and countries, by which we unify the same institution and country that has different appearances in our dataset, and also correct some inconsistency caused by typos.

## 3.2 Generating Co-author Networks

Co-author networks and two aggregated collaboration networks were created in CiteSpace[11]. In order to have component separated views of these network, which make identification of the largest components easier, we regenerated these networks in Pajek[3]. The network files were converted into an edgelist format for network analysis in NetworkX toolkit[4].

## 3.3 Network Analysis

The collaboration network analysis focuses on topological analysis, which employs various statistical measures to characterize the topology of collaboration networks[2].

- Network size: we report the number of nodes and edges. The network size shows the size of community of SDSS related study.

[4] network.lanl.gov

- Largest component (*LC*) size: A component is an isolated sub-network in a disconnected network. The largest component has the largest number of nodes among all components. We report the size of the LC with its number of nodes and edges in each network.

- Average distance (*L*): The average value of the shortest path length between any pair of nodes in the network. A shorter average distance means the collaboration between any pair of entities is closer. The average distance in a Erdos-Renyi model random network $L_{rand}$ with the number of *N* nodes and *M* edges is obtained using the following formula ([13] p. 144):
$L_{rand} = \ln(N)/\ln(2M/N)$

- Clustering coefficient (*C*): A network's clustering coefficient is the average clustering coefficient (*c*) over all the nodes, which is calculated as the ratio of the number of edges between the node's direct neighbors to the number of possible edges between the node's direct neighbors.

  *c*= (the number of edges between the neighbors)/(the possible number of edges between the neighbors)

  The clustering coefficient in a Erdos-Renyi model random network $C_{rand}$ with the number of *N* nodes and *M* edges could be obtained with the following formula[16]:

  $C_{rand} = 2M/N(N-1)$

From these topological measures, one can characterize and compare collaboration networks. Along with a time line, the dynamic patterns of collaboration can also be observed.

## 4. RESULTS

Collaboration networks and their topological measurements are presented in this section in the order of author-level collaboration, institutional collaboration and international collaboration networks. Then we report the comparison of the topological measurements across the three levels along with the dynamic pattern.

## 4.1 Collaboration Networks and Topological Measurements

### 4.1.1 Author level collaboration in SDSS

Figure 1 shows snapshots of the eight-year author level collaboration networks. Nodes are in red and edges are in grey scale. The eight year author collaboration networks are all dominated by one giant cluster. Besides the largest cluster, some authors formed relatively small clusters, and few authors worked alone, hence becoming single nodes without edges connected to the other nodes.

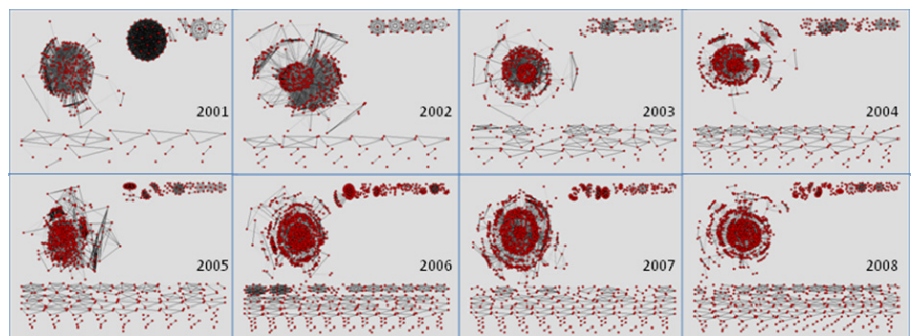

**Figure 1. Author collaboration networks in separated component view (2001-2008).**

**Table 1. Topological measurements of author collaboration networks**

| Year | $N$ | $E$ | $C$ | $C_{rand}$ | $L$ | $L_{rand}$ | LC-N (%) | LC-E (%) |
|---|---|---|---|---|---|---|---|---|
| 2001 | 324 | 7279 | **0.623** | **0.139** | 1.603 | 1.519 | 231(71.3) | 6583(90.4) |
| 2002 | 507 | 24407 | **0.828** | **0.190** | 1.777 | 1.364 | 425(83.8) | 24264(99.4) |
| 2003 | 616 | 21978 | **0.861** | **0.116** | 1.891 | 1.505 | 458(74.4) | 21641(98.5) |
| 2004 | 846 | 19702 | 0.779 | 0.055 | 2.332 | 1.755 | 641(75.8) | 19210(97.5) |
| 2005 | 1030 | 20088 | 0.817 | 0.038 | 2.629 | 1.894 | 743(72.1) | 19394(96.5) |
| 2006 | 1728 | 26945 | 0.694 | 0.018 | 2.624 | 2.167 | 1198(69.3) | 24397(90.5) |
| 2007 | 1533 | 23780 | 0.813 | 0.020 | 2.7 | 2.136 | 1125(73.4) | 22649(95.2) |
| 2008 | 1692 | 29028 | 0.766 | 0.020 | 2.81 | 2.103 | 1216(71.9) | 27924(96.2) |

$N$: the number of nodes; $E$: the number of edges; $C$: clustering coefficient; $C_{rand}$: clustering coefficient of a random network of the same size; $L$: average distance; $L_{rand}$: average distance of the random network; LC-N (%): the number of nodes in the largest component and the percentage to the total number of nodes; LC-E(%): the number of edges in the largest component and the percentage to the total number of edges.

Table 1 lists the topological measurements of these networks. The number of nodes in the SDSS collaboration network increased almost linearly throughout the eight years, while the number of edges, the co-author collaboration, has two drops at year of 2004 and 2007, but remains relatively constant.

In the early years (2001 to 2003), the values of clustering coefficient in author collaboration networks are relatively close to the random network at the same order of magnitude (highlighted in bold font), while in the later years the differences become larger at different order of magnitudes. The average distance of author collaboration network increased almost linearly in the eight years. The average distance is larger than the average distance of the random network, but still in the same order of magnitude.

As depicted in figure 1, a large number of nodes formed the largest component (around 70% to 80% of the total number of nodes), and these nodes forms the majority of collaborative relations (more than 90% of the total number of edges). These values show no clear trends in author collaboration network.

### 4.1.2 Institutional collaboration in SDSS
Figure 2 shows the snapshots of institutional collaboration networks over the eight years. The networks in Figure 2 also have a large dominant cluster with several much smaller clusters. The evolution of the institutional collaboration is analogous to a tree-like process, in which a few nodes in the middle of the largest cluster form a core; while the other nodes in this cluster expend from the core with few trunks, but there are very few interconnections among these peripheral nodes.

Table 2 lists the topological measurements of institutional collaboration networks in SDSS. The number of nodes and edges were increasing in the eight years except for a drop at 2007. Compared to a random network, the gaps in terms of the values of clustering coefficient of institutional collaboration networks were larger in 2001, then close in 2002 to 2004 in the same order of magnitude (highlighted in bold font), and became larger in the rest of the years. The values of clustering coefficient in institution networks were decreasing over time. The average distance in the institutional collaboration is smaller than the average distance in random networks for all years, and shows an increasing tendency.

The percentage of nodes and edges in the largest cluster to the total number of nodes and edges fluctuated in the eight years, showing no clear trends.

**Table 2. Topological measurements of institute collaboration networks**

| Year | $N$ | $E$ | $C$ | $C_{rand}$ | $L$ | $L_{rand}$ | LC-N (%) | LC-E (%) |
|---|---|---|---|---|---|---|---|---|
| 2001 | 46 | 41 | 0.22 | 0.040 | 2.402 | 6.623 | 28(60.9) | 33(80.5) |
| 2002 | 74 | 59 | **0.062** | **0.022** | 2.234 | 9.224 | 36(48.6) | 36(61.0) |
| 2003 | 104 | 110 | **0.079** | **0.021** | 3.137 | 6.199 | 72(69.2) | 92(83.6) |
| 2004 | 153 | 139 | **0.073** | **0.012** | 2.914 | 8.424 | 87(56.9) | 109(78.4) |
| 2005 | 190 | 163 | 0.074 | 0.009 | 2.807 | 9.719 | 96(50.5) | 118(72.4) |
| 2006 | 262 | 265 | 0.077 | 0.008 | 3.706 | 7.904 | 166(63.4) | 219(82.6) |
| 2007 | 260 | 252 | 0.032 | 0.007 | 3.383 | 8.401 | 156(60.0) | 205(81.3) |
| 2008 | 281 | 267 | 0.036 | 0.007 | 3.889 | 8.782 | 184(65.5) | 228(85.4) |

### 4.1.3 International collaboration in SDSS
Figure 3 shows the snapshots of the eight years international collaboration networks. There is also a major cluster in all networks. Visual observation reveals that the major component is a tree in 2001, and several core nodes formed the center of the networks in 2002 and 2003, and then the density of the networks increased in the more recent years. There are only a few nodes isolated from the major component, and the number of these nodes is less than ten.

Table 3 lists the topological measurements of the international collaboration networks. Similar to the institutional collaboration networks, the total number of nodes and edges in international collaboration networks has an increasing tendency except a drop at 2007. The values of clustering coefficient in country collaboration networks show the similar tendency as institution collaboration networks, which is close to random networks in the early years (highlighted in bold font) and became more different and much larger in the later years. The values of average distance in international collaboration networks are smaller than that of random

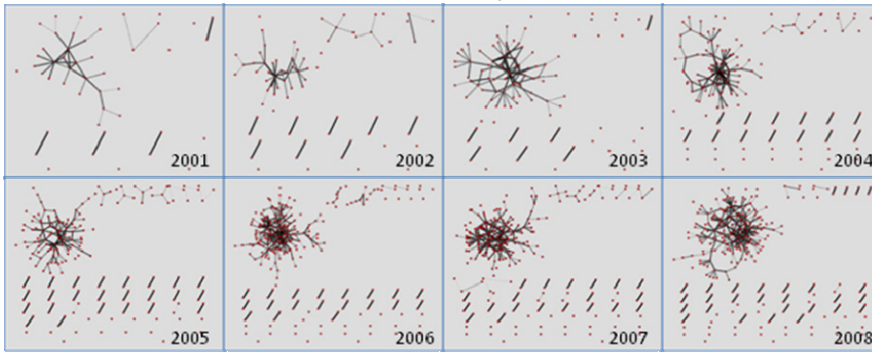

**Figure 2. Institute collaboration networks in separated component view (2001-2008).**

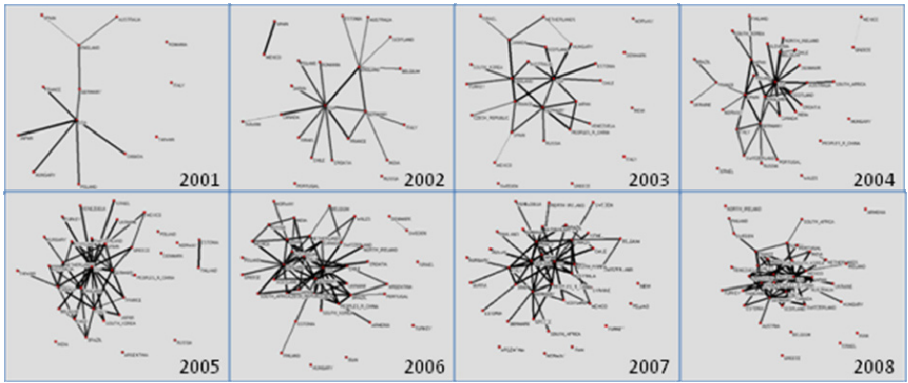

**Figure 3. International collaboration networks in separated component view (2001-**

**Table 3. Topological measurements of international collaboration networks**

| Year | $N$ | $E$ | $C$ | $C_{rand}$ | $L$ | $L_{rand}$ | LC-N (%) | LC-E (%) |
|------|-----|-----|-----|------------|-----|------------|----------|----------|
| 2001 | 13 | 11 | 0 | 0.141 | 2.035 | 6.623 | 11(84.6) | 10(90.9) |
| 2002 | 22 | 20 | **0.083** | **0.087** | 2.061 | 9.224 | 18(84.6) | 19(95.0) |
| 2003 | 27 | 30 | **0.083** | **0.085** | 2.061 | 6.199 | 18(81.8) | 29(96.7) |
| 2004 | 33 | 38 | 0.179 | 0.072 | 2.312 | 8.424 | 27(81.8) | 37(97.4) |
| 2005 | 33 | 52 | 0.23 | 0.098 | 1.746 | 9.719 | 25(75.8) | 49(94.2) |
| 2006 | 40 | 70 | 0.277 | 0.090 | 2.236 | 7.904 | 34(85.0) | 69(98.6) |
| 2007 | 38 | 58 | 0.281 | 0.083 | 2.31 | 8.401 | 32(84.2) | 58(100.0) |
| 2008 | 39 | 73 | 0.258 | 0.099 | 2.469 | 8.782 | 35(89.7) | 73(100.0) |

networks, and fluctuated in the eight years, showing no clear trends.

These largest components in international collaboration networks contain 80 to 90 percent of nodes and nearly all edges (95% to 100%).

## 4.2 Comparison of topological measurements across the three level collaborations

This section shows the comparison of the topological measurements across the three level collaboration networks.

Figure 4 depicts the network sizes at the three level collaboration networks over the eight years. Because the author level collaboration networks have a large number of nodes and edges than corresponding networks at the other two aggregated levels, we use logarithmic scale in the y axis. Figure 4 shows that the number of nodes at all three level collaboration networks increased in a similar pattern. The increasing number of nodes means the research community, including scientists, institutions and countries, of SDSS grew continuously. Surprisingly, the number of edges suggests a different scenario. While the numbers of edges in institutional and international collaboration networks increased with the similar trend as nodes did, the number of edges in author collaboration networks is nearly constant in the eight years. Therefore the average degree (the ratio of two times of the number of edges to the number of nodes) in author collaboration networks decrease, which means the average number of collaborators of a scientist decreased in SDSS collaboration.

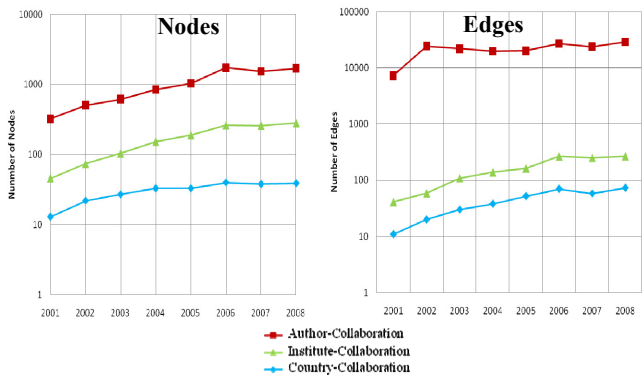

**Figure 4. Dynamic trends of network sizes in SDSS.**

Figure 5 shows the comparison of clustering coefficient across the three levels of collaboration networks over time. Author collaboration networks have larger values of clustering coefficient than corresponding networks at institutional and international levels, which is expected since the authors from the same institution tend to collaborate frequently than authors from other institutions and countries. International collaboration networks have larger clustering coefficient values than institutional collaboration. Because this study ignored the weight information of edges, a single publication co-authored between two countries' scientists will establish a collaboration link and be treated equivalently as links that represent many instances of collaboration. In this case, the collaboration in country level is expected to denser than institutional collaboration.

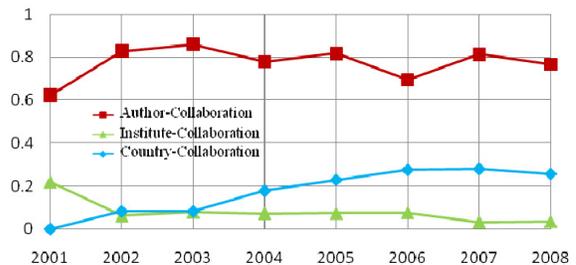

**Figure 5. Dynamic trends of clustering coefficient in SDSS.**

In terms of dynamic trends, the institutional collaboration networks show an interesting decrease, which means pairwise collaboration relations became popular, and two institutions that collaborated with the same third party institution is less likely to collaborate with each other. International collaboration has reverse tendency, as the values of clustering coefficient in country increased in the majority of years, two countries that collaborated with one common country become more likely to collaborate together.

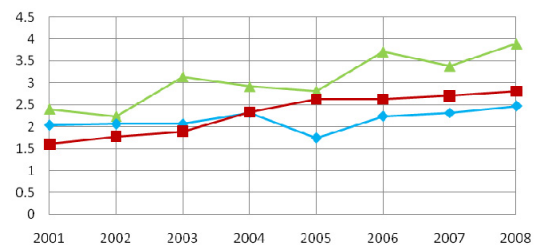

**Figure 6. Dynamic trends of average distance in SDSS.**

Figure 6 shows the trends of the values of average distance in three collaboration networks. The institutional collaboration networks have the longest average distance than the other two levels, and kept increasing while fluctuated several times. The longer average distance values mean that in general to establish a collaboration relation between two institutions is harder than find another person or country given the long intermediate institutions that needed to pass by. The average distance in author collaboration networks has an ascent tendency, from 1.5 to nearly 3, which means as the size of the SDSS community increased, finding another collaborator needs to pass more persons. The international collaboration has nearly the same average distance over the eight years, and is the smallest one.

The results in figure 5 and 6 together show an interesting pattern. While the clustering coefficient values in institutional collaboration networks kept decreasing, their average distance went up, suggesting that the institutional collaboration networks become sparser and more tree-like shape. This analysis result is consistent with direct observation in figure 2.

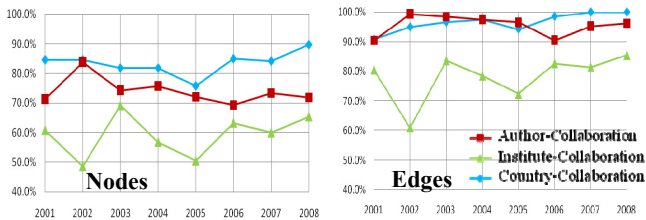

**Figure 7. Dynamic trends of proportions of the largest component to the entire network in SDSS.**

Figure 7 shows the proportions of the number of nodes and edges in the largest components to the total number of nodes and edges in the networks. In the three-level collaboration networks, Figure 7 shows no clear trends. Compared to other two levels, institutional collaboration has the lowest percentage of the number of nodes and edges in the largest components in all eight years. Nearly 40 percent of nodes and 20 percent of edges are outside the largest components. In author and international collaboration networks, the nodes isolated from the largest component are more likely to work along given the fact that they only count for less than 5% of the edges in the whole network. Figure 7 shows that in SDSS–related studies 60 percent of institutions dominate the majority collaboration relations, but still many institutions (around 40%) can carry on their own research by using SDSS data.

## 5. DISCUSSIONS

This section focused on the implications of the results in above section. Due to the paper length limit, we only highlight the most interesting and unexpected patterns derived from the results, including the random to small-world evolution patterns in all three level collaboration; the constant number of author level collaboration instances versus the increasing number of collaboration instances at the other two levels; and the tree-like network evolution process in institutional collaboration networks. We also discuss some limitations of this study.

First, in all three level collaboration networks, the clustering coefficient values all show a random to small-world evolving pattern. In the early years, the collaboration networks have clustering coefficient value in the same order of magnitude to the random networks of the same number of nodes and edges. In author level networks, the average distance in early years are also very close to the corresponding values in random networks. In later years, the collaboration networks all have larger clustering coefficients than random networks do. In the institutional and international collaboration, the average distance in collaboration networks is smaller than random networks, but still in the same order of magnitude.

The random to small-world evolution pattern may represent the actual process of the SDSS project. In the early year, a few scientists who were active members of the SDSS research community may randomly choose their collaborators to start research collaboration, and then when collaboration relationship grew up, the network became more like a small world network, where acquaintances to a common acquaintance become acquainted with each other. Examples may be shown as in [12], scientists tend to collaborate with Chilean institutions who have astronomical facilities that can help verify their discoveries from the SDSS dataset. The open data environment may facilitate the process since they have a common ground — the same SDSS data.

Second, even though the number of collaboration relations in author level varied, it shows a constant tendency, while at the other two levels, the number of collaboration relations increased over time. We may posit a plausible explanation for this phenomenon. In astronomy community, the phase of postdoctoral training is common, hence generating a circulation of young astronomers between various schools and groups. When postdocs moved around different institutions and countries every few years, they may still tightly collaborate with the same group of astronomers, but in effect increase the chance of institutional and international collaboration relationship. In this process, the SDSS data, which is widely accessible in different locations, may help to maintain the tight collaboration relationship among the postdocs and their collaborators. However, to answer whether open data is the only reason that can explain this phenomenon, future studies are needed with carefully designed comparisons of collaboration networks between open data and non-open data environments.

Third, the direct observation of network images and the topological measurements confirm the tree-like evolving process of the institutional collaboration networks. The decreasing values of clustering coefficient and increasing values of average distance in institutional collaboration networks suggested that in SDSS research the possibility of establishing a new collaboration relation between two institutions that have a common collaborator become lower and lower.

Why is the pairwise collaboration in institutional collaboration very common in the recent years? Does the open data trend have impacts on this phenomenon? On the one hand, according to Cumming and Kiesler[15], multi-university collaboration is hard, requiring good coordination skills and supportive mechanisms. In our results, the pairwise collaboration could be a better tradeoff of innovation opportunities versus coordination costs than collaborations involved with three or more institutions. On the other hand, the publically accessible SDSS data may even hinder the institutional collaboration. When data is widely available, competing for the first publication of a discovery could prevent different institutions from some collaborative engagements, like sharing methods and results, which are much likely and easier to take place within one institution.

Some limitations of this study possibly constrain the generality of the results.

First, this study describes the dynamic patterns in one open data eScience project without comparing to other open data projects, or non-open data studies. Hence explanations raised in this section could only be considered as exploratory impacts of open data on science collaboration. The discovered patterns, however, could lay down the foundation for further comparison studies.

Second, our study ignores the weight information in collaboration networks. All edges are considered equal no matter how many publications were co-authored by two nodes. This may biases the results, especially in higher aggregated level like country level. Further research could achieve more accurate results if it can take the weight information into consideration.

## 6. CONCLUSIONS

In this study we have investigated the dynamic collaboration patterns in an open data eScience environment via a case study of co-authorship relation in the SDSS publications. We studied the co-authorship collaboration networks at three levels, namely the author level, institutional level, and country level. By visualizing these collaborative networks and measuring their topological properties over time, we reached the following conclusions.

1) The collaboration networks in the SDSS community experienced an evolution from a random network to small-world networks. But the small world properties varied across the three different collaboration levels. The institutional collaboration has a much less like small-world ingredient, a tree-like evolving process.

2) In SDSS the number of collaboration relations at the author level is nearly constant, while at the institutional and country levels, the numbers were increasing. The open data trend could help to explain this observation, but future studies are needed to compare our results to other open data projects and non-open data projects.

3) Pairwise institutional collaboration became common in recent years in the SDSS community. The open access data may hinder the collaboration among multiple institutions.

To our humble knowledge, this is the first study focused on the collaboration patterns in an open data eScience environment. The methodology framework developed in this study can be used in other open data or non-open data domains, such as the iSchool community. More studies are eagerly needed to better understand the new open data trend in science and its impacts on science.

## 7. ACKNOWLEDGMENTS

This work is supported by the National Science Foundation under Grant No. IIS-0612129. Thomson Reuters provides the bibliographic data for the analysis. We also thank the anonymous reviewers for their constructive insights to the early draft. Funding for the SDSS and SDSS-II has been provided by the Alfred P. Sloan Foundation, the Participating Institutions, the National Science Foundation, the U.S. Department of Energy, the National Aeronautics and Space Administration, the Japanese Monbukagakusho, the Max Planck Society, and the Higher Education Funding Council for England. The SDSS Web Site is http://www.sdss.org/.